\documentclass[twocolumn,showpacs,amsmath,amssymb,prl]{revtex4}

\usepackage{graphicx}

\begin{document}

\title
{Negative differential thermal resistance and  thermal transistor}

\author{Baowen Li}

\affiliation{Department of Physics and Center for Computational
Science and Engineering, National University of Singapore,
Singapore 117542, Republic of Singapore
\\
Laboratory of Modern Acoustics and Institute of Acoustics, Nanjing
University, 210093, PR China
\\
NUS Graduate School for Integrative Sciences and Engineering,
Singapore 117597, Republic of Singapore }
\author{Lei Wang}
\affiliation{ Department of Physics and Center for Computational
Science and Engineering, National University of Singapore,
Singapore 117542, Republic of Singapore}
\author{Giulio Casati}
\affiliation{Department of Physics and Center for Computational
Science and Engineering, National University of Singapore,
Singapore 117542, Republic of Singapore\\
 International Center for
Nonlinear and Complex Systems,
Universita' degli studi dell'Insubria, Como, Italy,\\
Istituto Nazionale di Fisica della Materia, Unita' di Como, and
 Istituto Nazionale di Fisica Nucleare, sezione di Milano,
Milano, Italy}

\date{12 December 2005, revised 3 March 2006, to appear in Appl. Phys. Lett.}

\begin{abstract}
We report on the first model of a thermal transistor to control
heat flow. Like its electronic counterpart, our thermal transistor
is a three-terminal device with the important feature that the current
through the two terminals can be controlled by small changes in the temperature
 or in the current through the third terminal. This control feature allows us to
 switch the device between ``off" (insulating) and ``on" (conducting) states or
 to amplify a small current.  The thermal transistor model is possible because
 of the negative differential thermal resistance.
\end{abstract}


\maketitle

The invention of the transistor\cite{Bardeen} and other relevant
devices that control the electric charge flow has  led to an
impressive technological development. In recent years some
interesting progress has been made also in the control of the heat
current. Indeed a theoretical model of a thermal rectifier has
been recently proposed\cite{rectifier} in which the heat can flow
preferentially in one direction. More recently,  new models of
thermal diode have been devised\cite{diode1,diode2} which allows
to improve the efficiency by more than three orders of magnitude.

In this Letter  we propose a model for a thermal transistor which
can control the heat flow in analogy to the usual electric
transistor for the control of the electric current. The reason why
such a device can, in principle, operate is grounded in the
phenomenon of ``{\it negative differential thermal resistance}''
(NDTR)which, as shown below, can take place in nonlinear lattices.

\begin{figure}[ht]
\includegraphics[width=\columnwidth]{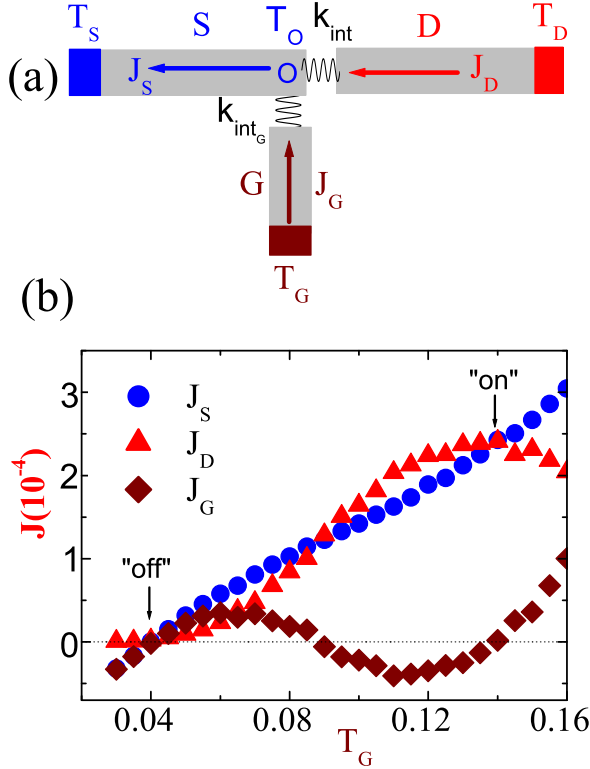}
\vspace{-1cm} \caption{\label{fig:switch}(a) Configuration of the
thermal transistor. (b) Heat current versus the control
temperature $T_G$. Parameters are: $T_D$=$0.2$, $V_D$=$1.0$,
$k_D$=$0.2$, $k_{int}$=$0.05$; $T_S$=$0.04$, $V_S$=$5$,
$k_S$=$0.2$, $V_G$=$5$, $k_G$=$1$, $k_{int_G}$=$1$. Notice that in
a wide region both $J_S$ and $J_D$ increase when the temperature
$T_G$ is increased. }
\end{figure}

Our model of thermal transistor consists of three segments, D, S
and G. (see Figure 1(a) for its configuration).  The names D, S
and G follow from the ones used in a MOSFET
(Metal-Oxide-Semiconductor Field-Effect-Transistor).  Each segment
is
 a Frenkel-Kontorova (FK) lattice
\cite{Fkreview0,Fkreview1}. Segments S and D are coupled (usually
weakly) via a spring of constant $k_{int}$, while Segment G, the
control segment, is coupled to the junction particle O between
segments S and D via a spring of constant $k_{int_G}$. Temperature
$T_G$ is used to control temperature $T_O$ (at the junction O) which
determines the heat current from D to S. The total Hamiltonian of
the model writes:
\begin{align} \label{HAM}
H&=H_S+H_D+H_G+H_{int}+H_{int_G} \nonumber \\
 &=\sum_{i=1}^{N_S}\frac{1}{2}{\dot{x}^2_{S,i}}+\frac{1}{2}k_S(x_{S,i}-x_{S,i-1})^2-\frac{V_S}{(2\pi)^2}\cos 2\pi x_{S,i} \nonumber \\
 &+\sum_{i=1}^{N_D}\frac{1}{2}{\dot{x}^2_{D,i}}+\frac{1}{2}k_D(x_{D,i}-x_{D,i-1})^2-\frac{V_D}{(2\pi)^2}\cos 2\pi x_{D,i} \nonumber \\
 &+\sum_{i=1}^{N_G}\frac{1}{2}{\dot{x}^2_{G,i}}+\frac{1}{2}k_G(x_{G,i}-x_{G,i-1})^2-\frac{V_G}{(2\pi)^2}\cos 2\pi x_{G,i} \nonumber \\
 &+\frac{1}{2}k_{int}(x_{S,N_S}-x_{D,N_D})^2 \nonumber \\
 &+\frac{1}{2}k_{int_G}(x_{S,N_S}-x_{G,N_G})^2
\end{align}

 Here $x_{D,i}$, $x_{S,i}$, and $x_{G,i}$ are the particles' displacements
 from their equilibrium positions in
segment $D$, $S$ and $G$, respectively. Fixed boundaries conditions are
taken, i.e., $x_{D,0}=x_{S,0}=x_{G,0}=0$.

In the following we will present first the results of our numerical
simulations, then the theoretical explanation. In our numerical
simulations the lattices are coupled, at their first particles, with heat baths
at different temperatures $T_D$, $T_S$ and $T_G$, respectively.
We use Langevin heat baths and we have
checked that our results do not depend on the particular heat bath
realization (e.g. Nose-Hoover heat baths). The local heat flux is
defined by $J_n\equiv k\langle \dot{x}_n(x_n-x_{n-1})\rangle$ and
the local temperature is defined as $T_n=\langle
\dot{x}_n^2\rangle$. The simulations are performed long enough to
allow the system to reach a non-equilibrium steady state where the
local heat flux is constant in each segment.

 As illustrated in Figures 1-2, the model described by Hamiltonian (\ref{HAM})
 can exhibit different useful functions depending on the parameters values.

\begin{figure}[ht]
\includegraphics[width=\columnwidth]{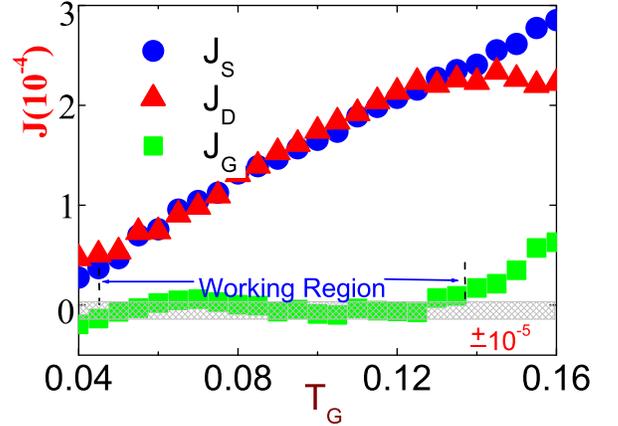}
\vspace{-1cm} \caption{\label{fig:ampl} Heat current versus the
control  temperature $T_G$. Here: $T_D$=$0.2$,  $V_D$=$1.0$,
$k_D$=$0.2$, $T_S$=$0.04$, $V_S$=$5$, $k_S$=$0.2$,
$k_{int}$=$0.05$, $V_G$=$5$, $k_G$=$1$, $k_{int_G}$=$0.1$. The
shadow region is the range of variation of $J_G$ in the
temperature interval $T_G$$\in(0.05, 0.135)$. }
\end{figure}

\emph{Thermal Switch}: We first demonstrate the ``\textit{switch}"
function of our transistor, namely we show that the system can act
like a good heat conductor or an insulator depending on the
control temperature  $T_G$. This is shown in Figure 1(b), where we
plot $J_G, J_S$, and $J_D$ versus $T_G$.  When $T_G$ increases
from $0.03$ to $0.135$, both $J_D$ and $J_S$ increase. In
particular, at three points: $T_G$$\approx$ $0.04$, $0.09$ and
$0.135$, we have that $J_D$=$J_S$  and thus $J_G$ is exactly zero.
These three points correspond to ``off", ``semi-on" and ``on"
states, at which $J_D$ is $2.4\times 10^{-6}, 1.2\times 10^{-4} $
and $2.3\times 10^{-4}$, respectively. The ratio of the heat
currents at the ``on" and ``off" states is about 100,  hence our
model displays one important function - switch - just like the
function of a MOSFET used in a digital circuit.

\emph{Thermal modulator/amplifier}: As demonstrated above, the
heat current from $D$ to $S$ can be switched between different
values. However, in many cases, like in an analog circuit, one
needs to continuously adjust the current in a wide range. In
Figure 2 we demonstrate this ``modulator/amplifier" function of
our transistor. Here it is seen that in the temperature interval
$T_G$$\in$ $(0.05, 0.135)$, the heat current through the segment G
remains very small ($-10^{-5}\sim10^{-5}$), within the shadow
strip in Figure 2, while the heat currents $J_S$ and $J_D$ are
continuously controlled from $5\times 10^{-5}$ to $2\times
10^{-4}$.

Let us now turn to the discussion of the underlying physical
mechanism which allows the transistors function. In all cases
considered here, the heat resistance of segment G has been taken
small enough so that $T_O\approx T_G$. For fixed thermal baths
temperatures $T_S$ and $T_D$, we would like now to understand how
the heat currents $J_S$ and $J_D$ depend on temperature $T_O$. This
dependence is determined by the differential thermal resistance,
$R_S=\left(\frac{\partial J_S}{\partial
T_O}\right)^{-1}_{T_S=\mbox{const}}$ and $R_D=-\left(\frac{\partial
J_D}{\partial T_O}\right)^{-1}_{T_D=\mbox{const}}$. In typical
situations  the heat current increases with the increase of the
thermal gradient and therefore both $R_S$ and $R_D$ are positive. As
a consequence, the ``{\it current amplification factor}",
 \begin{equation}
\alpha \equiv \left|\frac{\partial J_D}{\partial
J_G}\right|=\left|\frac{R_S}{R_S+R_D}\right| <1,
\end{equation}
namely, in order to induce a change $\Delta J_D$, the control  heat
bath has to provide a larger $\Delta J_G$. This means that the
``transistor'' can never work!

\begin{figure}[ht]
\includegraphics[width=\columnwidth]{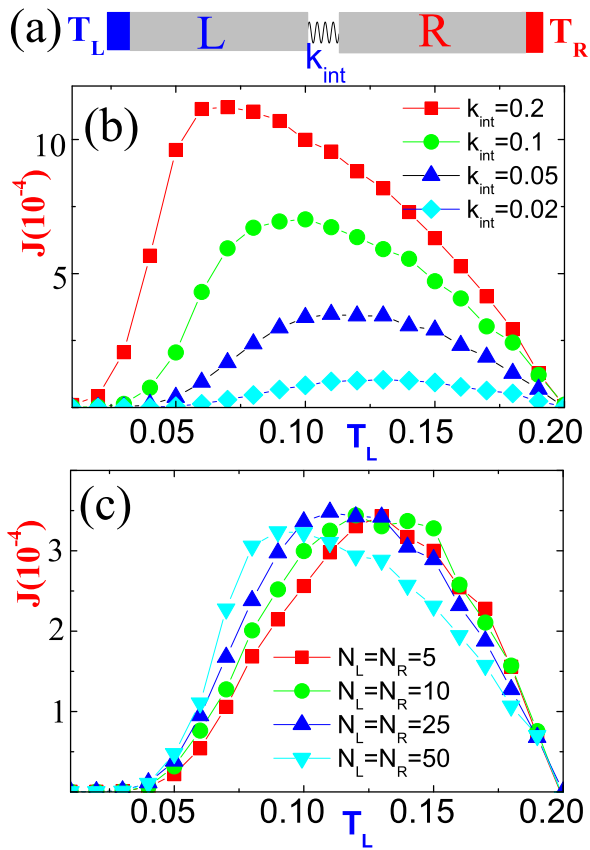}
\vspace{-1cm} \caption{\label{fig:ndtr}(a) Configuration of the
 model exhibiting the negative differential thermal resistance (see
definition in the text). (b) Heat current versus temperature $T_L$
(at fixed $T_R$=$0.2$) for different coupling constants,
$k_{int}$, with lattice size $N_L$=$N_R$=$25$. The system
parameters are: $V_L$=$5$, $V_R$=$1$, $k_L$=$1$, $k_R$=$0.2$. (c)
Same as (b) but for different system sizes and $k_{int}$=$0.05$.
Notice that when $T_L$$\le$$0.1$ the heat current increases with
decreasing the external temperature difference. }
\end{figure}

In order to get the amplification effect, namely $\alpha>1$,
either $R_S$ or $R_D$ must be negative.  Thus we must have a
``{\it negative differential thermal resistance}'' (NDTR). In such
case the possibility arises that when $T_O$ changes both $J_S$ and
$J_D$ change simultaneously in the same way. Therefore the
condition $J_S$=$J_D$ can be achieved for several different values
of $T_O$ (as shown in Figure 1). Analogously, the condition
$J_S\approx J_D$ can be verified in a wide range of $T_O$ (as
shown in Figure 2). In these situations heat switch and heat
modulator/amplifier are possible. In the ideal, limiting case of
$R_S$=$-R_D$ which, in principle, can be obtained by adjusting
parameters, the transistor works perfectly.

Figure 3 illustrates the mechanism of NDTR. In Figure 3(b) we plot
the heat current versus temperature $T_L$ (see Figure 3(a))for
fixed $T_R$, fixed lattice size and for different coupling
constants $k_{int}$ .  In Figure 3(c) instead the heat current is
plotted for different lattice sizes with fixed coupling constants
$k_{int}$.

As it is clearly seen, in a wide range of system parameters, for a
fixed temperature $T_R$, there exists a temperature interval in
which a larger value of $T_L (<T_R)$, corresponding to a smaller
thermal gradient, can induce a larger heat current.

The phenomenon of NDTR can be understood from the mismatch between
the phonon bands of the two interface particles. This is
illustrated in Figure 4 where we plot the power spectra of the two
interface particles at different temperatures $T_L$ for a fixed
$T_R$=$0.2$  (Figure 3(a)). As it is seen, the overlap of the
power spectra increases as $T_L$ increases i.e. as the thermal
gradient decreases. Therefore, by increasing $T_L$ we are in the
presence of two competing effects: from one hand the thermal
gradient decreases; on the other hand the band overlap increases.
The beaviour of the heat current depends on which of these two
effects prevails. In particular the phenomenon of NDTR can take
place.

The physical mechanism of such behaviour is due to the fact that
the variation of the bands position with temperature depends on
the strength of nonlinearity. This general property is at the base
of the thermal diode models \cite{diode1,diode2}. In particular,
for the model discussed in this paper, as the temperature $T_L$
increases, the kinetic energy of the left interface particle
becomes large enough to overcome the on site potential barrier and
low frequency modes appear.

Finally, we would like to discuss the possibility of building a
thermal transistor in real physical systems and the possible
applications. Due to the fast developing semiconductor technology,
it is now possible to fabricate several low dimensional devices,
such as nanotubes, nanowires and thin films in nanometer
scale\cite{Cahill03}. A nanotube with small diameter or a nanowire
is basically  a one dimensional heat conductor. For the purpose of
our model, which is indeed one dimensional, one can fabricate a
T-shape nanotube or nanowire. To build the on-side potential in
the FK model, one may put different segments of
nanotubes/nanowires on the surface of different substrates.

\begin{figure}[ht]
\includegraphics[width=\columnwidth]{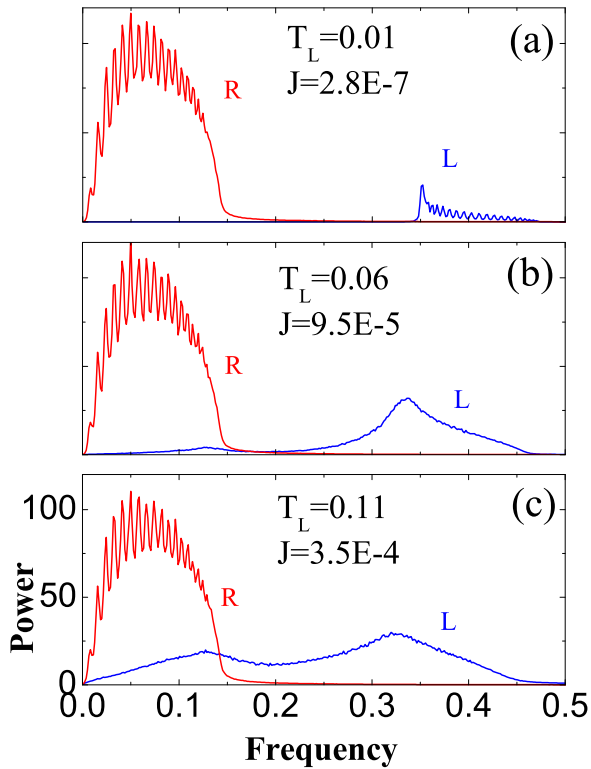}
 \caption{\label{fig:spectra}The power spectra of
the particles at the right (R) side and the left (L) side of the
interface, for different temperature $T_L$ with fixed $T_R$=$0.2$.
The system parameters are: $V_L$=$5$, $V_R$=$1$, $k_L$=$1$,
$k_R$=$0.2$, $k_{int}$=$0.05$, $N$=$50$. Notice that the overlap
of the spectra increases as the temperature $T_L$ increases, which
results in the increase of the heat current with decreasing the
external temperature difference when $T_L$$\le $$0.11$.}
\end{figure}

Among the possible applications we would like to mention, as an
example, that in our model, by adjusting the temperature of the
third terminal, one can amplify the current through the other two
terminals. This can be used as an efficient heat pump to dissipate
heat from a microelectronic or nanoelectronic device where a large
amount of redundant heat is produced due to the fast operational
speed of the device. This could be very important since, as it is
known, the redundant heat, if it is not dissipated immediately,
may deteriorate the device.

In conclusion, we have built a theoretical model for a thermal
transistor. The model displays two basic functions of a
transistor: switch and modulator/amplifier. The crucial element of
the thermal transistor is the \emph{negative differential thermal
resistance}. Although at present this is no more than an abstract
model, we believe that, sooner or later, it can be realized in a
nanoscale system experiment. After all the Frenkel-Kontorova model
used in our simulations is a very popular model in condensed
matter physics\cite{Fkreview0,Fkreview1}. For instance, it has
been used to model crystal dislocations, epitaxial monolayers on
the crystal surface, ionic conductors and glassy materials, an
electron in a quasi-1D metal below the Peierls transition, charge
density waves, Josephson junctions chains, and dry friction. More
recently, this model has been employed to study transport
properties of vortices in easy flow channels\cite{Fkvotex} and
strain-mediated interaction of vacancy lines in a pseudomorphic
adsorbate system\cite{Fkvacancy}.
\bigskip

This project is supported in part by FRG of NUS and the DSTA
Singapore under Project Agreement POD0410553 (BL). LW is supported
by DSTA Singapore under Project Agreement POD0001821. GC is also
partially supported by EU Contract No. HPRN-CT-2000-0156 (QTRANS)
and by MURST (Prin 2003, Ordine e caos nei sistemi estesi non
lineari: strutture, stocasticita' debole e trasporto anomalo).

\end{document}